\documentclass[journal]{IEEEtran}

\usepackage{graphicx}

\begin{document}

\title{HEER: Hybrid Energy Efficient Reactive\\ Protocol for Wireless Sensor Networks}

\author{N. Javaid$^{\ddag}$, S. N. Mohammad$^{\ddag}$, K. Latif$^{\ddag}$, U. Qasim$^{\pounds}$, Z. A. Khan$^{\$}$, M. A. Khan$^{\ddag}$\\\vspace{0.4cm}
        $^{\ddag}$COMSATS Institute of Information Technology, Islamabad, Pakistan. \\
        $^{\$}$Faculty of Engineering, Dalhousie University, Halifax, Canada.\\
        $^{\pounds}$University of Alberta, Alberta, Canada.
     }

\maketitle

\begin{abstract}
Wireless Sensor Networks (WSNs) consist of numerous sensors which send sensed data to base station. Energy conservation is an important issue for sensor nodes as they have limited power. Many routing protocols have been proposed earlier for energy efficiency of both homogeneous and heterogeneous environments. We can prolong our stability and network lifetime by reducing our energy consumption. In this research paper, we propose a protocol designed for the characteristics of a reactive homogeneous WSNs, HEER (Hybrid Energy Efficient Reactive) protocol. In HEER, Cluster Head(CH) selection is based on the ratio of residual energy of node and average energy of network. Moreover, to conserve more energy, we introduce Hard Threshold (HT) and Soft Threshold (ST). Finally, simulations show that our protocol has not only prolonged the network lifetime but also significantly increased stability period.
\end{abstract}


\section{Related Work}
Heinzelman \emph{et al}. [1], introduced the first hierarchical clustering algorithm for WSNs, called LEACH. It is one of the most popular protocols in WSNs. The main idea is to form clusters of the sensor nodes. LEACH outperforms classical clustering algorithm by using adaptive clustering and rotating CHs. This saves energy as transmission will only be performed on that specific CH rather than all the nodes. LEACH performs well in homogeneous environment however, it's performance deteriorates in heterogeneous environment.

Threshold-sensitive Energy Efficient Network (TEEN) [2] is a reactive protocol for time critical applications. The CH selection and cluster formation of nodes is same as that of LEACH. In this scheme, CH broadcasts two threshold values i.e. Hard Threshold (HT) and Soft Threshold (ST). HT is the absolute value of an attribute to trigger a sensor node. HT allows nodes to transmit the event, if the event occurs in the range of interest. Therefore, this not only reduces transmission to significant numbers but also increases network lifetime.

Georgios \emph{et al}. [3], proposed a two level heterogeneous aware protocol, consisting of normal and advance (high energy) nodes. It is based on the weighted election probabilities of each node according to their respective energy to become a CH. Intuitively, advance nodes have more probability to become a CH than normal nodes, which seems logical according to their energy consumption. Stable Election Protocol (SEP) does not require any global knowledge of the network. The drawback of SEP is that it does not consider the changing residual energy of the node hence, the probability of advanced nodes to become CH remains high irrespective of the residual energy left in the node. Moreover, SEP performs below par if the network is more than two levels.

In [4], authors proposed Distributed Energy Efficient Clustering (DEEC) protocol for WSNs. DEEC is a clustering protocol for two and multilevel heterogeneous networks. The probability for a node to become CH is based on residual energy of the nodes and average energy of network. The epoch for nodes to become CHs is set according to the residual energy of a node and average energy of the network. The node with higher initial and residual energy has more chances to become a CH than the low energy node.

\section{Motivation}
A number of routing protocols have been proposed in the area of WSNs. Most of them are based on CH selection. However, not much attention has  been devoted towards time critical applications. Most of the routing protocols are for proactive networks. DEEC, being a proactive heterogeneous network protocol is not well suited for time critical applications. TEEN is a reactive protocol which guarantees that in homogeneous environment, unstable region will be short. After the death of the first node, all the remaining nodes are expected to die on average within a small number of rounds as a consequence of the uniform remaining energy due to the well distributed energy consumption. On the other hand TEEN in the presence of high energy nodes yields a large unstable region. The reason being, all high energy nodes are equipped with almost the same energy however, the CH selection process is unstable and as a result most of the time these nodes are idle, as there is no CH to transmit. Hence, in our research paper we focus on developing a protocol that gives us better results for time critical applications in both environments i.e. homogeneous and heterogeneous environment.

\section{Performance Measuring Parameters}
We define following parameters which evaluate and compare the performance of clustering protocols.

\subsection{Stability Period}
It is the time interval when the network starts its operation till the death of the first node. It is also referred to as stable region or steady state.

\subsection{Instability Period}
It is the time interval from the death of the first node until the death of the last node or till the time the network is dead. It is referred as unstable region.

\subsection{Network life time}
It is the time interval from the start of operation of the network till the death of last alive node:
             $Network life time=Stability Period+Instability Period$

\subsection{Alive Nodes}
This instantaneous measure shows the total number of nodes (advanced, normal) alive i.e. the nodes having energy greater than zero.

\subsection{Throughput}
It is the total rate of data sent over the network from nodes to their respective CHs and from CHs to base station.

\section{Protocols in WSNs}
In this section, we describe the functionality and characteristics of our model of a WSNs. We particularly present the setting, the CH selection and how transmissions occur.

\subsection{TEEN}
TEEN is the first reactive protocol. In this scheme, closer nodes form clusters with a CH to transmit the collected data to one upper layer. This is same as LEACH protocol however, at every cluster change time, the CH broadcasts two threshold values i.e hard and ST. HT is the absolute value of an attribute to trigger on its transmitter and report to its respective CH. HT allows nodes to transmit data, if the data occurs in the range of interest. Therefore, a significant reduction of the transmission delay occurs. Moreover, ST is the small change in the value of the sensed attribute. Next transmission occurs when there is a small change in the sensed attribute once it reaches the HT. So, it further reduces the number of transmissions.

\subsection{DEEC}
DEEC is a proactive protocol designed for two and multi level heterogeneous networks.  All the nodes use the initial and residual energies to select a CH.  The node with higher initial and residual energy has greater probability to become a CH.
In a two-level heterogeneous network, we have two categories of nodes, m.N advanced nodes with initial energy equal to $E_{0}.(1+a)$ and (1-m).N normal nodes, where  $E_{0}$ is the initial energy. Moreover, $a$ and $m$ are two variables which control the nodes (advanced or normal) percentage types and $E_{total}$ is the total energy in the network. The value of $E_{total}$ is given as:

\begin{equation}
    E_{total} = N(1-m) E_{0}+Nm E_{0}(1+a)=N E_{0}(1+am)
\end{equation}

The value of the total initial energy of the multi-level heterogeneous networks is given as:

\begin{equation}
    E_{total}= E_{0}(N+\sum_{i=1}^{N}a_i )
\end{equation}

The probability of a  node to become CH  in a two level heterogeneous network is:

\begin{equation}
    P_i=P_{opt} E _i (r) /(1+am)\bar{E}(r)
\end{equation}

For normal nodes

\begin{equation}
    P_i=P_{opt} E _i (r) /(1+am)\bar{E}(r)
\end{equation}

For advanced nodes this model can be easily extended to multi-level heterogeneous networks:

\begin{equation}
    P (s_i) =P_{opt}N (1+a_i)/ (N+a_i)
\end{equation}

As we are assuming uniformly distributed node, so distance of cluster members from CH is:

\begin{equation}
    d_{toCH} =M/\sqrt{2\pi k}
\end{equation}

Average distance between base station and CH is:

\begin{equation}
    d_{toBS}=0.765M/2
\end{equation}

\begin{figure}[t]
\centering
\includegraphics[height=8cm,width=7cm]{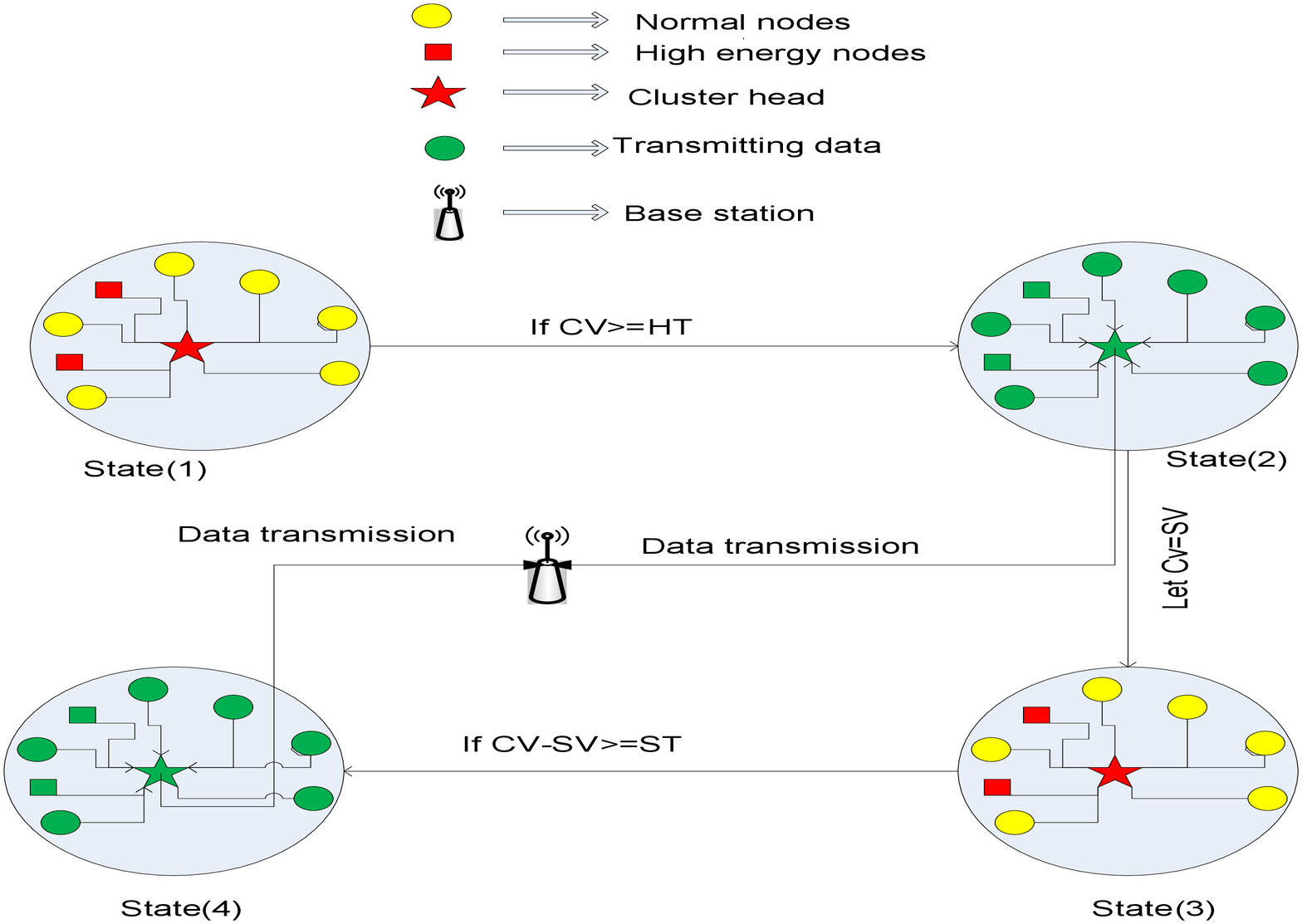}
\caption{HEER from data sensing to data transmission for a cluster}
\end{figure}

\subsection{HEER Protocol}
In this section, we describe HEER, which improves the stable region for clustering hierarchy process for a reactive network in homogeneous and heterogeneous environment. We use the initial and residual energies of the nodes to become CH similar to that of DEEC.  It does not require any global knowledge of energy at any election round.When cluster formation is done, the CH transmits two threshold values, i.e $HT$ and $ST$. The nodes sense their environment repeatedly and if a parameter from the attributes set reaches its $HT$ value, the node switches on its transmitter and transmits data. The Current Value ($CV$) on which first transmission occurs, is stored in an internal variable in the node called Sensed Value ($SV$). This reduces the  number of transmissions.
Now the nodes will again transmit the data in same cluster period when $CV-SV \ge ST$. That is, if $CV$ differs from $SV$ by an amount equal to or greater than $ST$, then it further reduces the number of transmissions.

Figure. 1 shows different states of a cluster i.e. from data sensing  to data transmitting. Every node selects itself as a CH on the basis of its initial energy and residual energy. In State (1)  a cluster is formed the node senses its environment continuously until the parameter (CV) reaches its HT value. When CV reaches HT value, the nodes become green as shown in the figure in State (2). The node then switches on its transmitter and sends the data to the CH. The CH aggregates and transmits data to base station. The CV on which first transmission occurs is stored in SV. The node, then again starts sensing its environment as shown in State (3) until the CV differs from SV by an amount equal to or greater than ST.  When this condition becomes true, the node again switches on its transmitter and sends data to CH.The CH then transmits data to base station as shown in State (4) of figure. 1.

\subsubsection{Important Features}

\begin{itemize}
\item HEER performs best for time critical applications in both homogeneous and heterogeneous environment.
\item It reduces the number of transmissions resulting in the reduction of energy consumption.
\item It increase the stability period and network lifetime.
\end{itemize}

\section{Throughput in Proactive and Reactive Protocols}
Proactive protocols sense their environment and transmit data periodically.  They consume energy continuously with time due to periodic transmission. Our main focus in proactive protocols is on increasing lifetime, throughput and to decrease energy consumption. Contrary to proactive protocol, reactive protocol is application dependent. It senses the environment periodically but transmits data only when its $CV$ reaches to absolute value of the attribute. As data transmission consumes more energy than data sensing, so, in reactive network the throughput can be minimized or maximized as per its application. The throughput in reactive networks is inversely proportional to the network lifetime or its stability period. If transmissions are less the stability period and network lifetime will be prolonged as $CV$ does not reach the absolute value . However, if the $CV$ reaches $HT$ value (absolute value) repeatedly then maximum number of transmissions will occur and nodes will die quickly.

\section{Simulation Results}
In this section, we simulate an environment with varying temperature in different regions. Our field has dimensions of  $100 \times 100$ square units. The number of nodes in the field is $n=100$. We assume that base station is in the center of sensing nodes. To evaluate the performance of HEER, we simulate it with TEEN and DEEC. The parameters used in our simulation are listed in table 1. Our goals in conducting the simulation are as follows.

\begin{itemize}
\item{We examine the performance of TEEN and HEER for the prolonging of stability period and network life time.}
\item{We also observe the throughput of both the protocols.}
\end{itemize}

We observe the performance of TEEN,DEEC and HEER under homogeneous environments. We also examine the sensitivity of our protocol to the degree of heterogeneity in the sensor network.

\begin{table}
\caption{Parameters used in our simulations}
    \begin{tabular}{|l|l|}
        \hline
         \textbf{Parameter}	                                        & \textbf{Value}           \\ \hline
        Initial energy, $E_0$                               &  0.5 J           \\ \hline
         Transmitting and Receiving energy, $E_{elect}$     & 5nJ/bit          \\ \hline
        Amplification energy for short distance, $E_{fs}$   & 10pJ/bit/m2     \\  \hline
        Amplification energy for long distance, $E_{mp}$    & 0.013 pJ/bit/m2 \\  \hline
        Energy for data aggregation, $E_{DA}$               & 5nj/bit/signal   \\ \hline
        Percentage of advanced nodes, m                     & 0.1              \\ \hline
         Energy of advance nodes                            & $E_{0}$(1+a)         \\
        \hline
    \end{tabular}
\end{table}

\begin{figure}[t]
\centering
\includegraphics[height=7cm,width=9cm]{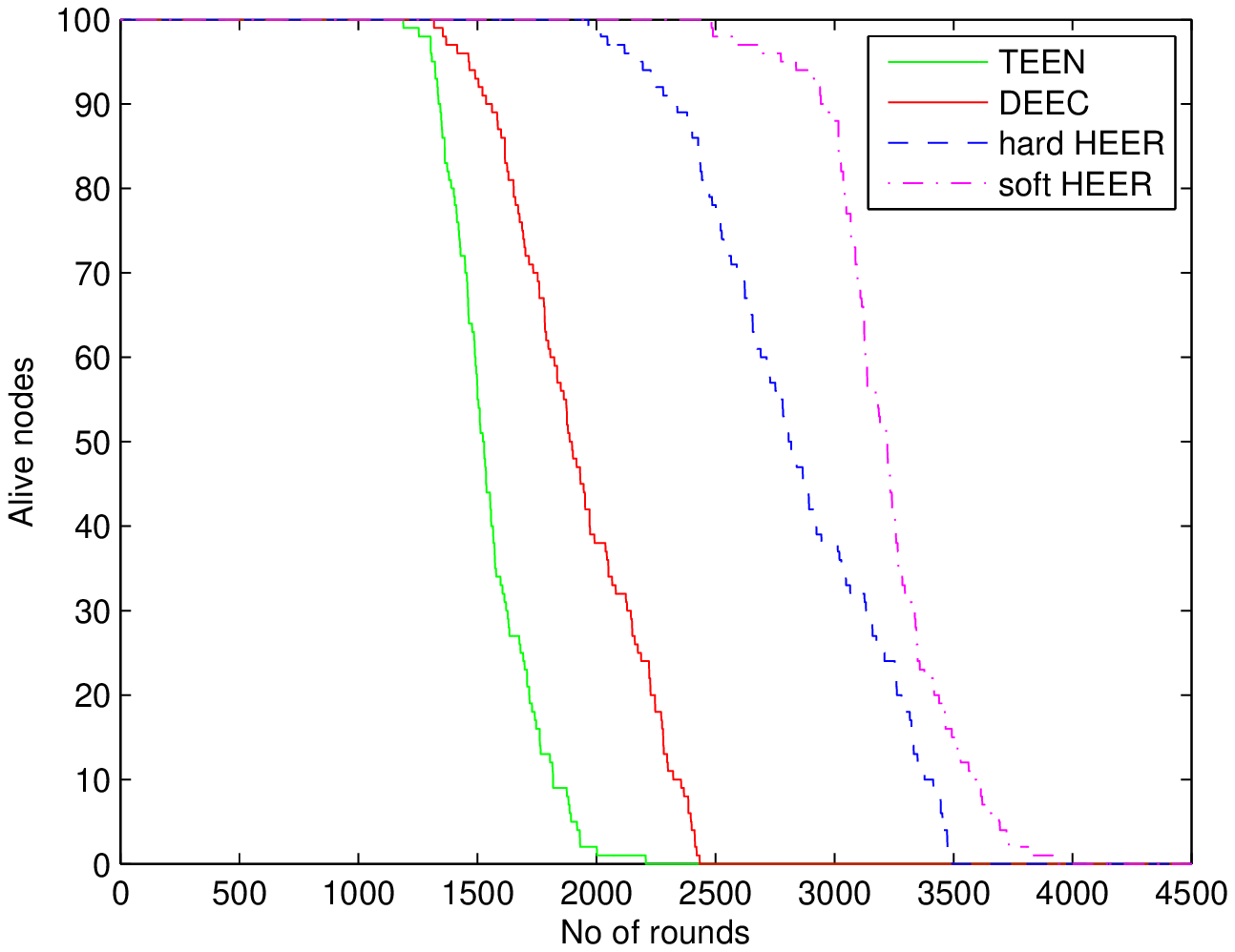}
\caption{Homogeneous environment: when $HT$=100 and $ST$=2}
\end{figure}

\begin{table}
    \caption{Comparison table: when $HT$=100 and $ST$=2}
    \renewcommand{\tabcolsep}{0.1cm}
    \begin{tabular}{|l|l|l|l|l|}
        \hline
        \textbf{Protocol}  & \textbf{Stability Period} & \textbf{Life Time}         & \textbf{Environment}  &  \textbf{Classification} \\ \hline
        Teen      & 1221             & 1947              & Homogeneous  & Reactive       \\ \hline
        DEEC      & 1395             & 2461              & Homogeneous  & Proactive      \\ \hline
        hard HEER & 2005             & 3595              & Homogeneous  & Reactive       \\ \hline
        soft HEER & 2493             & 3959              & Homogeneous  & Reactive       \\
        \hline
    \end{tabular}
\end{table}

\begin{figure}[t]
\centering
\includegraphics[height=7cm,width=9cm]{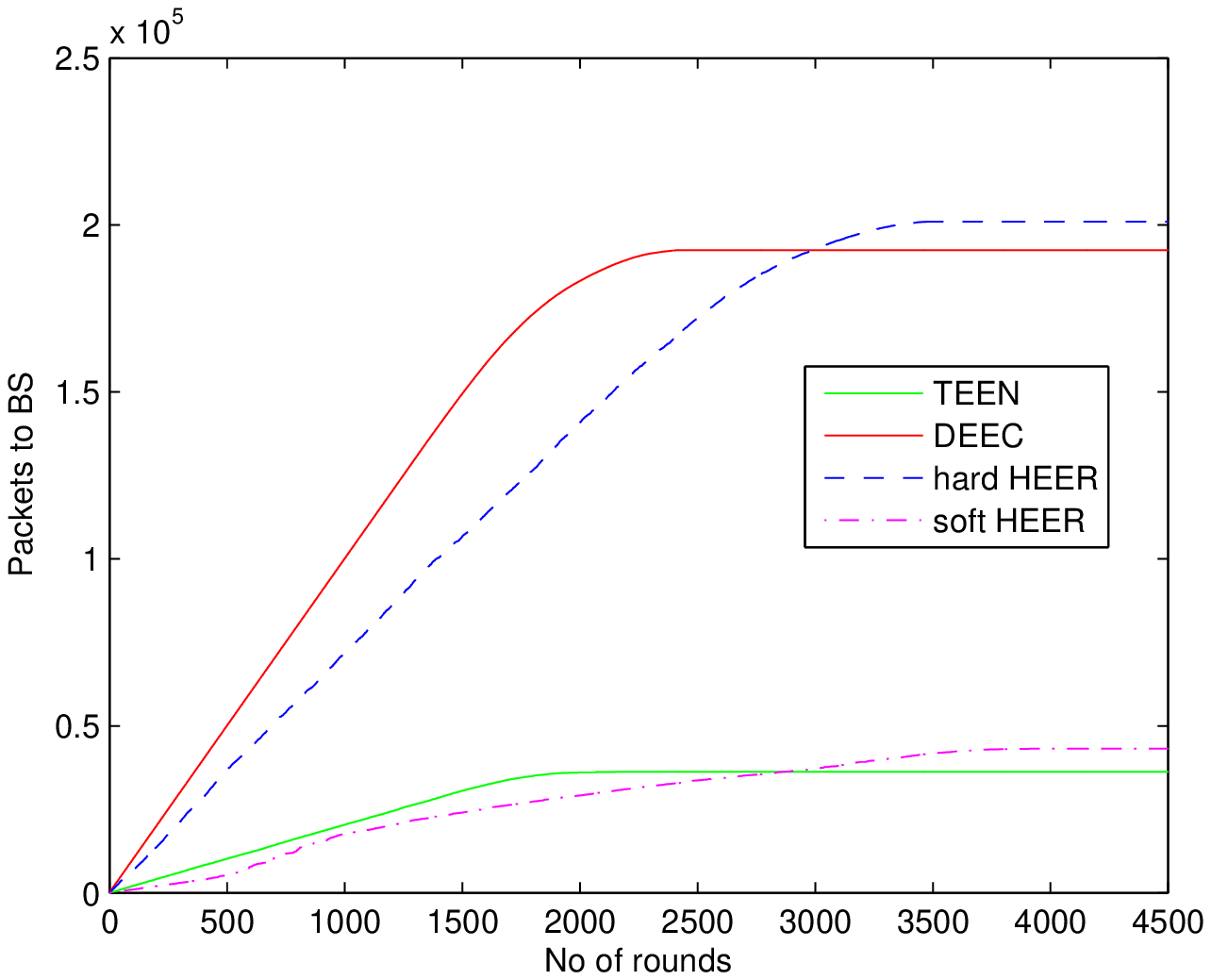}
\caption{Homogeneous environment: when $HT$=100 and $ST$=2}
\end{figure}

\begin{figure}[t]
\centering
\includegraphics[height=7cm,width=9cm]{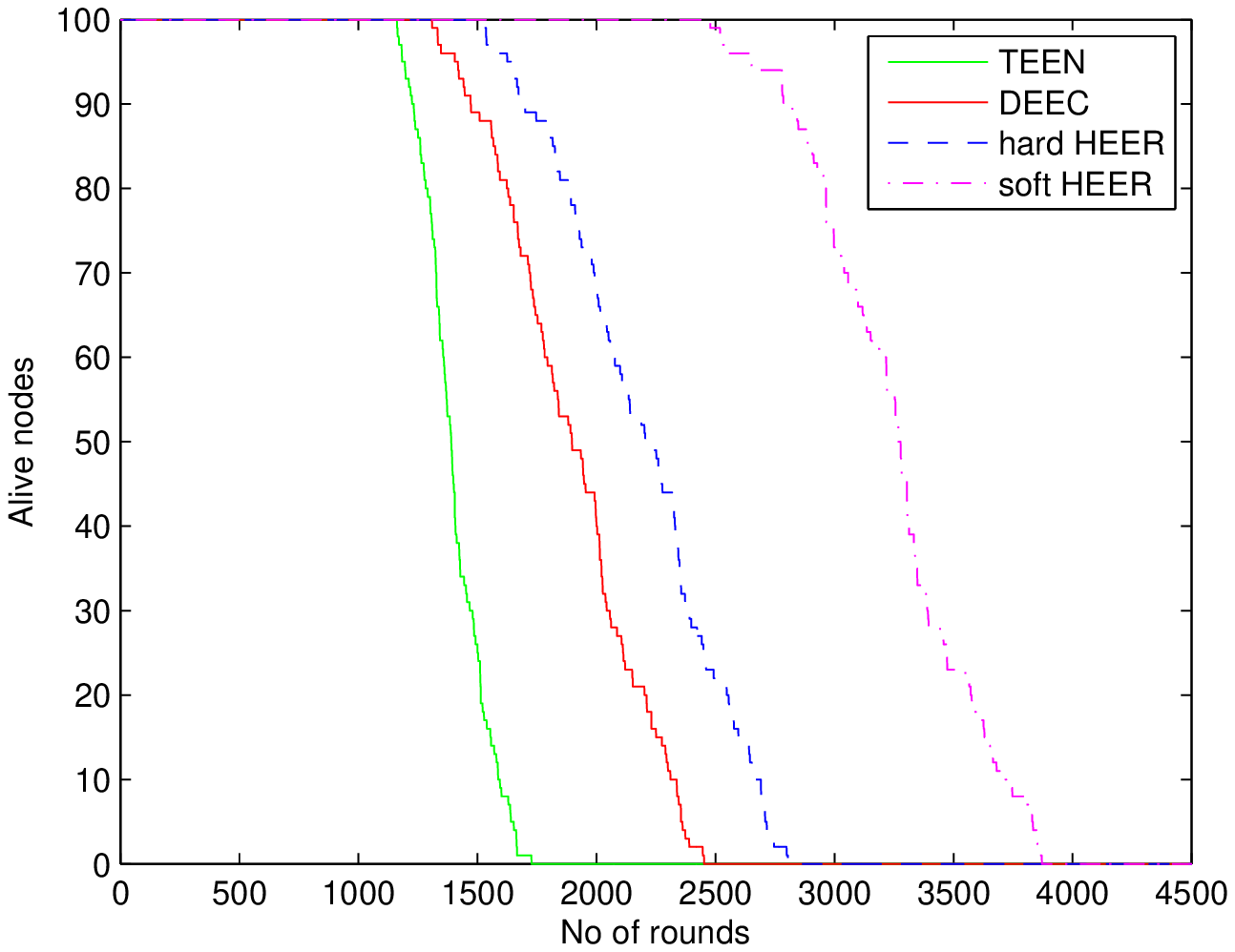}
\caption{Homogeneous environment: when $HT$=70 and $ST$=10}
\end{figure}

\begin{figure}[t]
\centering
\includegraphics[height=7cm,width=9cm]{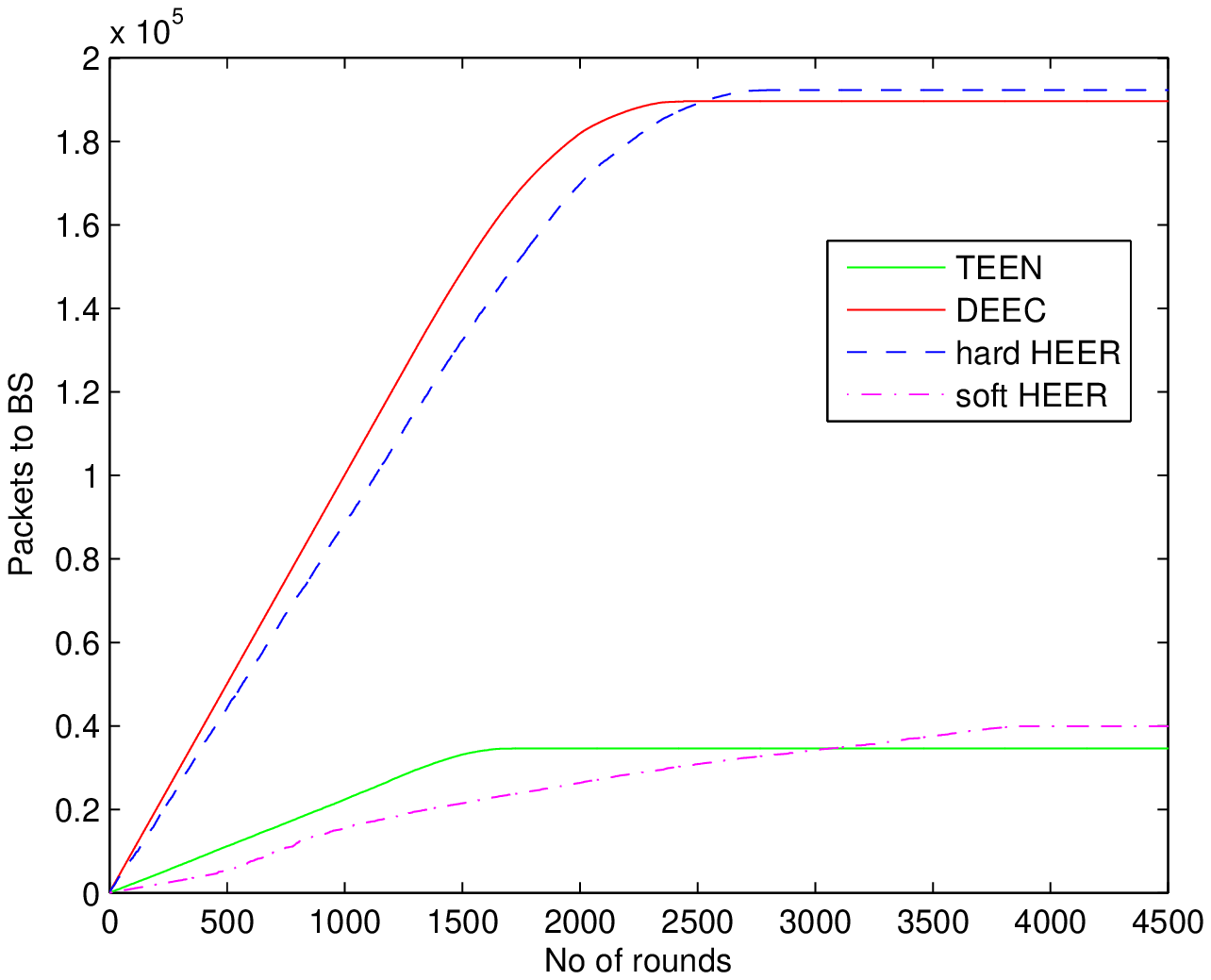}
\caption{Homogeneous environment: when $HT$=70 and $ST$=10}
\end{figure}

\subsubsection{Homogeneous Environment}
In this section, we compare HEER, TEEN and DEEC protocols in homogeneous environment. We observe from figure. 3 that in TEEN after the death of first node, all the remaining nodes die within a small number of rounds. This is due to the reason that all the nodes have same probability to become a CH.
DEEC prolongs the stability period and network life time due to the election of CH on the basis of residual energy. High residual energy nodes have greater probability of becoming a CH.

The stability period of HEER ( hard and soft) is much longer than that of TEEN and DEEC. HEER introduces HT value which decreases the number of transmissions to base station. This increases the stability period and network lifetime. ST further reduces the number of transmissions resulting in the reduction of energy consumption. This prolongs the network life time. HEER outperforms DEEC in terms of stability period and network life time by a factor of 1.78 and 1.60 respectively. Moreover, both stability period and network life time of HEER also outperforms TEEN by a factor of 2.0. From figure. 5, we can observe that by changing the values of thresholds, the stability period and network life time of HEER (hard) changes noticeably. The stability period and network lifetime decreases if we decrease the value of $h$ as shown in figure. 5. Moreover, figure.6 shows that the throughput difference between HEER (hard) and DEEC decreases by decreasing the difference between the two thresholds. ST also effects the network lifetime. If number of transmissions increase, we observe a decrease in network lifetime and vice versa.

\section{Conclusion}
In this paper, we present a hybrid reactive protocol of TEEN and DEEC for homogeneous environment. HEER minimizes the energy consumption by first distributing load to all high energy nodes and then on to low energy nodes. Like TEEN, it is well suited for time critical applications and is more efficient than TEEN and DEEC.

\end{document}